\documentclass[epj]{svjour}
\usepackage{graphics}

\begin{document}

\title{Interfacial adsorption in Potts models on the square lattice}

\author{N.G. Fytas,\inst{1} A. Malakis,\inst{1,2} W. Selke,\inst{3} \and
L.N. Shchur\inst{4}}

\institute{ \inst{1} Applied Mathematics Research Centre, Coventry
University, Coventry, CV1 5FB, United Kingdom \\\inst{2}
Department of Physics, Section of Solid State Physics, University
of Athens, Panepistimiopolis, GR 15784 Zografou, Greece \\
\inst{3} Institut f\"{u}r Theoretische Physik and JARA-HPC, RWTH
Aachen University, 52056 Aachen, Germany \\ \inst{4} Landau
Institute for Theoretical Physics, 142432 Chernogolovka, Russia}

\date{Received: date / Revised version: date}

\abstract{We study the effect of interfacial phenomena in
two-dimensional perfect and random (or disordered) $q$-state Potts models with
continuous phase transitions, using, mainly, Monte Carlo
techniques. In particular, for the total interfacial adsorption,
the critical behavior, including corrections to scaling, are
analyzed. The role of randomness is scrutinized. Results are
discussed applying scaling arguments and invoking findings for
bulk critical properties. In all studied cases, i.e., $q = 3$,
$4$, and $q = 8$, the spread of the interfacial adsorption
profiles is observed to increase linearly with the lattice size at
the bulk transition point.
\PACS{
      {75.10.Hk}{Classical spin models}   \and
      {05.50+q}{Lattice theory and statistics (Ising, Potts. etc.)}  \and
      {05.10.Ln}{Monte Carlo method, statistical theory}
     }
}
\authorrunning{N.G. Fytas et al.} \titlerunning{Interfacial adsorption in
 Potts models on the square lattice}

\maketitle

\section{Introduction}
\label{sec:intro}

At the interface between coexisting phases various intriguing
phenomena may occur, like roughening and wetting~\cite{Abra,Diet}.
An interesting complication arises when one considers the
possibility of more than two phases. A third phase may be formed
at the interface between the two other phases. An experimental
realization is the two-component fluid system in equilibrium with
its vapor phase~\cite{Diet}. The situation may be mimicked in a
simplified fashion via multi-state models by fixing distinct
boundary states at the opposite sides of the system. The formation
of the third phase, with an excess of the non-boundary states, in
such models has been called ``interfacial
adsorption''~\cite{Huse,Fish}.

Various aspects of interfacial adsorption, including its novel
critical behavior at the bulk phase transition and the interplay
with wetting, have been investigated quite extensively. In
particular, specific models, like Potts and Blume-Capel models,
have been studied using numerical techniques, especially, Monte
Carlo (MC) methods and density renormalization-group calculations
~\cite{Huse,Yeo,Kroll,Yama,Carlon,Alba,Fytas,Alba2,Alba3}.
Furthermore, scaling and analytic arguments have been presented
~\cite{Huse,Kroll,Carlon,Lebo,Messa,Cardy,Delf}. In particular,
critical exponents and scaling properties of the temperature and
lattice size dependencies have been determined. The fundamental
role of the type of the bulk transition has been clarified, with
isotropic scaling holding at continuous and tricritical bulk
transitions, while interfacial adsorption is described by
anisotropic scaling at a bulk transition of first-order type.

Despite the resulting interesting insights, open questions still
exist. In this article, we shall mainly consider two aspects which
have been largely neglected before, the adsorption profiles and
the effect of randomness on the interfacial adsorption in
two-dimensional (2D) Potts models \cite{Wu}. Concretely, we shall
consider the $q$-state Potts model on the square lattice, where
the ferromagnetic nearest neighbor couplings may take the values
$J_1$ or $J_2$, occurring with the same probability and being
distributed randomly. Of course, when the ratio $r=J_2/J_1$ takes
the value $1$, one deals with the pure (or perfect) case. In the
dilute (or random) case, the position of the interface, as well as
the extent of the intervening third phase of non-boundary states,
may be strongly affected by the spatial distribution of the
couplings. Because of the inhomogeneity of a given bond
realization, it seems, obviously, worthwhile to monitor not only
the total excess adsorption of the non-boundary states but also
the local structure of the interfacial adsorption, especially the
adsorption profiles.

The bulk transition temperatures of these Potts models are known
exactly from self-duality for arbitrary values of the internal
states $q$ and disorder-strength ratios $r$~\cite{Kinzel}.
Accordingly, analyses on the critical behavior of the interfacial
adsorption, based on extensive MC simulation data, as it is also
done in the present paper, are significantly simplified.

Bulk criticality of such random Potts models on the square lattice
has attracted much interest, partly, because the transition is of
continuous type for all values of $q$, while being, in the perfect
case, of first order for $q>4$~\cite{Wu,Berche}. Then, the
analysis of the interfacial adsorption in these models may be
simplified by the fact, that isotropic finite-size scaling is
expected to hold at continuous transitions~\cite{Huse,Kroll,Yama}.
Static and dynamic bulk critical properties of the random Potts
models have been estimated, supposedly, rather accurately, using a
variety of, predominantly, numerical methods~\cite{Berche}. Some
of the these results will turn out to be very useful in our study
on the interfacial adsorption.

Attention should be drawn to related previous work on interfacial
phenomena in dilute ferromagnetic Potts models, in particular,
considering hierarchical lattices, i.e. applying the
Migdal-Kadanoff real space renormalization to the square
lattice~\cite{Monthus}, or performing a preliminary MC study for
the square lattice model~\cite{Brener}.

The outline of the article is as follows: In the next
Section~\ref{sec:mm}, the model and the methods, especially, MC
simulations of Metropolis and Wolff type, will be introduced,
followed by the discussion of our main results in
Section~\ref{sec:results}. The summary, Section~\ref{sec:summary},
will conclude the article.

\section{Model and methods}
\label{sec:mm}

We shall study the nearest-neighbor random-bond $q$-state Potts
model on the square lattice described by the Hamiltonian
\begin{equation}
\label{eq:Hamiltonian} \mathcal{H}=-\sum_{\langle (i,j)\rangle}J_{i,j}\delta_{s_i,s_j},
\end{equation}
where the Potts variable at site $i$, $s_{i}$, takes the values
$1,2,\ldots,q$~\cite{Wu}. The ferromagnetic random couplings
$J_{i,j}>0$ between nearest neighbor sites $i$ and $j$ are either,
with probability $p$, $J_1$ or, with probability $1-p$, $J_2$. In
the case $J_1 > J_2$, one has either strong or weak bonds. Of
course, $r=J_2/J_1=1$ denotes the pure (or perfect) $q$-state
Potts model. In our study, we shall consider Potts models with $q
= 3$, $4$, and $q = 8$.

In this article, we shall consider the system at its self-dual
point, where both couplings occur with the same probability,
$p=1/2$. Then the phase transition temperature, $k_{\rm B} T_{\rm
c}/J_1$, between the ordered ferromagnetic phase and the
high-temperature disordered phase is known to follow
from~\cite{Kinzel}
\begin{equation}
\label{eq:Critical point} \left(e^{(J_1/k_{\rm B} T_{\rm
c})}-1\right) \left(e^{(rJ_1/k_{\rm B}T_{\rm c})}-1 \right) = q.
\end{equation}

In the random case, the phase transition is of continuous type for
all values of $q$, while, in the perfect case, the transition is
continuous only for $q \le 4$, being of first order at larger
number of Potts states $q$. Exact values of the critical exponents
are, so far, only known in the perfect case~\cite{Wu}. Numerical
analyzes, in the dilute case suggest that the bulk critical
exponents depend rather mildly on $q$~\cite{Berche}.

\begin{figure}
\resizebox{1 \columnwidth}{!}{\includegraphics{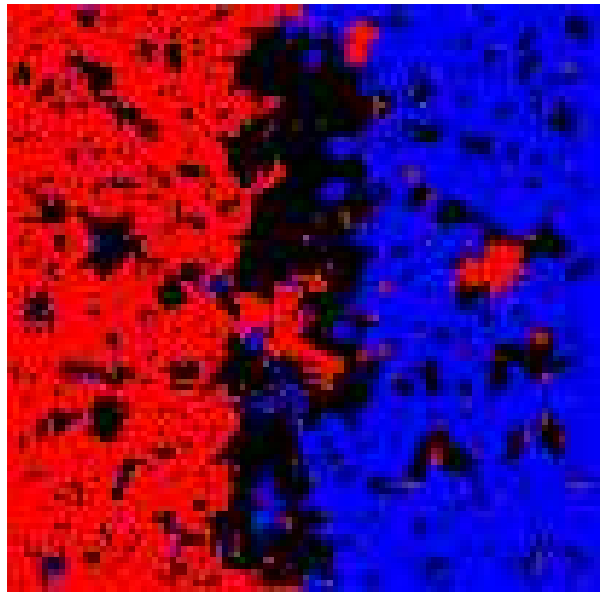}}
\caption{(color online) Typical equilibrium MC configuration of
the $L = 100$ $q = 8$ random-bond ($r = 1 / 10$) Potts model at
temperature $T = 0.98 T_{c}$ showing an interface for a particular
bond-disorder realization. Red color depicts the $q=1$ states,
blue color the $q = 2$ states, whereas the non-boundary states ($q
\ge 3$) adsorbed at the interface are shown blackened. Note that
the fixed boundaries $[1:2]$ are also included in this
illustration.} \label{fig1}
\end{figure}

The degeneracy between the $q$ equivalent Potts states may be
lifted by appropriate boundary conditions. Here, to study
interfacial adsorption, we shall employ special boundary
conditions, distinguishing the cases $[1:1]$ and $[1:2]$. For the
case $[1:1]$, the Potts variable is set, at all boundary sites,
equal to $q=1$, while for the case $[1:2]$, the variable is set
equal to $1$ at one half of the boundary sites, and to $2$ at the
opposite half of the boundary sites. Then, the boundary condition
$[1:2]$ introduces an interface between the 1-rich domain (or
phase) and the 2-rich domain (or phase). By examining typical MC
equilibrium configurations, as shown in Figure~\ref{fig1}, it is
seen that at the interface between the 1- and and 2-rich domains
an excess of the non-boundary states is generated compared to the
case in the absence of an interface.

Then, the interfacial adsorption, $W$, measuring the surplus of
non-boundary states induced by the interface between the 1-rich
and 2-rich regions, is defined, for lattices with $L^{2}$
non-boundary sites, with $L$ being the linear dimension of the
lattice, by~\cite{Huse}
\begin{equation}
\label{eq:ia} W=\frac{1}{L} \sum_n \sum_i
\left[(\delta_{s_i,n})_{[1:2]}-(\delta_{s_i,n})_{[1:1]}\right],
\end{equation}
summing over all non-boundary sites $i$ and over all non-boundary
states $n=3,4,\ldots,q$. The thermal average is taken. Obviously, $W$ 
may be interpreted as the
effective width of the domain of non-boundary states between the
1- and 2-rich domains.

Another useful quantity we shall consider is the profile, $w_{l}$,
of the interfacial adsorption, measuring the surplus of the
non-boundary states in line, $l$, parallel to the ideal, straight
interface. Then,
\begin{equation}
\label{eq:ia} W = \sum_l w_{l}.
\end{equation}
One may expect $w_l$ to fall off as one moves away from the
interface. Its maximum indicates the location of the interface,
which is, for a single realization of the Potts model with random
couplings, not necessarily in the center of lattice. The spread,
$d_{\rm w}$, of the adsorption profile may be measured by the
distance, at which $w_l$ decreases to half of its maximal value,
using a linear interpolation for the adsorption between successive
lines, as will be discussed below.

Using MC techniques, we recorded, in addition to the interfacial
properties, standard thermodynamic quantities, for both types of
boundary conditions. In particular, we measured the thermally
averaged energy, $E_{1:1}$ and $E_{1:2}$, the specific heat given
by the energy fluctuations, $C_{1:1}$ and $C_{1:2}$, and the order
parameter given by the majority fraction of the Potts
states~\cite{Wu,Berche}, $m_{1:1}$ and $m_{1:2}$.

In our simulations of the Potts models on square lattices with
$L^2$ sites, we applied the Metropolis and the cluster-flip Wolff
algorithm~\cite{LanBin}. Of course, cluster flips violating the
boundary conditions are not allowed~\cite{Gamsa}. As usual, small
lattices may be simulated using the Metropolis algorithm, while
the Wolff algorithm is more efficient and is preferred for larger,
say $L > 30$, system sizes. Overall, we studied lattices with up
to $100^2$ sites for the random Potts model and $200^2$ sites for
its pure counterpart.

Certainly, equilibration and averaging times depend on the lattice
size. Moreover, for random models, we observed that the given bond
realization may affect these times. In case of the Metropolis
algorithm, eventually, simulations with $10^7$ Monte Carlo steps
per site for $L=10$ were performed, increasing the length of the
runs, roughly, with $L^2$. In case of the Wolff algorithm, the
number of clusters $\mathcal{R}_{\rm cl}$ used in our simulations
varied from $2\times 10^7$ for the smaller systems sizes up to
$3\times 10^9$ for the larger sizes considered (see also
Table~\ref{tab:1}).

\begin{table}
\caption{Numerical details used in the Wolff simulations of the
8-state ($r=1/10$) random-bond Potts model on the square lattice.
In particular, the first column marks the linear size of the
lattice, the second column the number of clusters
$\mathcal{R}_{\rm cl}$ (created during averaging) and the third
the number of the independent disorder realizations
$\mathcal{N}$.} \label{tab:1}
\begin{tabular}{ccc}
\hline\hline\noalign{\smallskip} $L$ & Number of clusters $\mathcal{R}_{\rm cl}$ &  Disorder samples $\mathcal{N}$  \\
\noalign{\smallskip}\hline\noalign{\smallskip}
10   & $2\times 10^7$ & $20000$\\
20   & $2\times 10^7$ & $5000$\\
30   & $2\times 10^7$ & $4000$\\
40   & $2\times 10^7$ & $5000$\\
50   & $30\times 10^7$ & $4200$\\
60   & $40\times 10^7$ & $3200$\\
70   & $50\times 10^7$ & $1000$\\
80   & $70\times 10^7$ & $1000$\\
90   & $200\times 10^7$ & $1000$\\
100  & $300\times 10^7$ & $1000$\\
\noalign{\smallskip}\hline\hline
\end{tabular}
\end{table}

The main source of error is, for random systems, the fact that the
simulation data may vary quite drastically from bond configuration
to bond configuration. The corresponding histograms or
distributions have been recorded, especially, for the 8-state
Potts model with $r=1/10$. Bulk properties of this model have been
studied quite extensively
before~\cite{Berche,Chen,Jacob,Chat,Pala,Picco}. Indeed, in the
present study, we focused much attention on this case as well. The
histograms at the critical point, for the various quantities
discussed above, show nearly Gaussian shapes, but being weakly
tailed, in accordance with previous observations and discussions
for dilute Potts models on hierarchical lattices~\cite{Monthus}.
The standard errors resulting from an ensemble average over bond
realizations decrease with the number of configurations,
$\mathcal{N}$, approximately, proportionally to
$1/\sqrt{\mathcal{N}}$. The proportionality factor seems to become
somewhat smaller for larger lattices. To obtain reasonable
accuracy, as will be elucidated below, we averaged over a large
number of different bond configurations, as given explicitly in
Table~\ref{tab:1}. For pure Potts models ($r=1$) error bars follow
from averaging over a few MC runs employing different random
numbers, as usual. Typically, somewhat shorter runs are needed
compared to those for the random models. Usually, the error bars
are not depicted in the figures, because they are smaller than the
sizes of the symbols.

To determine critical properties from the MC data, we use
finite-size scaling arguments. For example, for the interfacial
adsorption, $W$, one expects~\cite{Huse}
\begin{equation}
\label{eq:scal} W \approx  L^{a}\Omega(t L^{1/\nu}),
\end{equation}
with $a = 1-\beta/\nu$ and the reduced critical temperature
$t=|T-T_{\rm c}|/T_{\rm c}$. $\Omega$ is the scaling function;
$\beta$ and $\nu$ are the usual bulk critical exponents for the
order parameter and the correlation length. A more refined ansatz
invokes corrections to the asymptotic scaling behavior, as will be
discussed in the following Section.

\section{Results}
\label{sec:results}

\subsection{Interfacial properties}
\label{sec:inad}

\begin{figure}
\resizebox{1 \columnwidth}{!}{\includegraphics{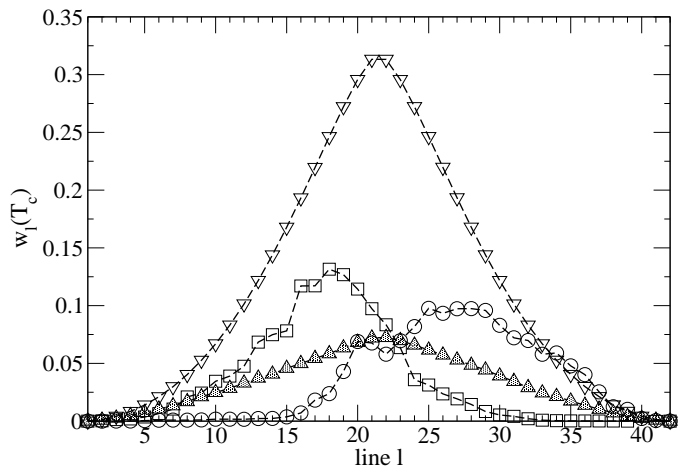}}
\caption{Critical adsorption profiles $w_l$ in the dilute 8-state
Potts model, $r=1/10$, for a single random-bond configuration
(circles and squares) and for an average over $\mathcal{N} = 120$
realizations (triangles up). For comparison, the critical
adsorption profile in the pure 8-state Potts model is depicted
(triangles down). Square lattices with $40^2$ sites have been
simulated.} \label{fig2}
\end{figure}

In the perfect 2D $q$-state Potts models the total interfacial
adsorption $W$ is known to vanish at zero and infinite
temperature, with a maximum, for finite lattices, near the
critical temperature $T_{\rm c}$. In the thermodynamic limit,
$W(T,L)$ diverges with characteristic critical
exponents~\cite{Huse,Carlon,Lebo}. In the present article, we
shall take a closer look at the interfacial adsorption by
analyzing the corresponding profile, $w_l$, as well. Moreover, the
impact of randomness on the interfacial properties will be
studied.

\begin{figure}
\resizebox{1 \columnwidth}{!}{\includegraphics{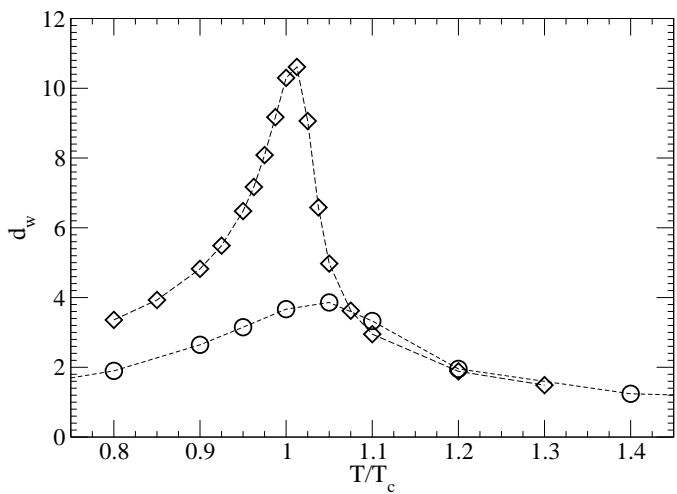}}
\caption{Temperature dependence of the spread of the adsorption
profiles $d_{\rm w}(T)$ in the perfect 3-state Potts model, as
observed in simulations of lattices with $20^2$ (circles) and
$60^2$ (diamonds) sites.} \label{fig3}
\end{figure}

Indeed, due to randomness in the couplings, the geometry of the
interface between the 1- and 2-domains may be rather complicated.
Straightforward considerations on the ground-state energy show
readily that, in contrast to the perfect case, the interface for a
given random-bond configuration is at sufficiently small values of
$r$, not necessarily straight or in the center of the lattice.
Depending on that configuration, there may be even overhangs,
i.e., interfaces may not satisfy the SOS criterion. As a
consequence, the adsorption profile $w_l$, for a fixed bond
configuration, may, at non-zero temperatures, display a maximum
away from the center of the lattice, as illustrated in
Figure~\ref{fig2}. As shown in that figure as well, after
averaging over the ensemble of bond realizations, $w_{l}$ is
expected to be symmetric about the center of the lattice.

The total interfacial adsorption $W$, at the critical point, for
lattices of small and moderate sizes, tends to increase with the
number of non-boundary states, $q-2$, and it is smaller in the
random case compared to the perfect model. This latter observation
may be explained by the fact that in the random case the system
is, both for predominantly strong and weak bonds, locally
effectively below or above criticality, while the interfacial
adsorption is expected to be maximal close to $T_{\rm
c}$~\cite{Huse,Carlon}.

Figure~\ref{fig2} indicates that the spread in the adsorption
profiles, $d_{\rm w}$, as defined in the previous Section, seems
to increase fairly weakly due to randomness. It is found to depend
rather mildly on $q$. Note that the temperature dependent spread
$d_{\rm w}(T)$ displays a maximum close to the critical point, as
illustrated in Figure~\ref{fig3}, similar to the behavior of
$W$~\cite{Huse,Carlon}. Its critical behavior, in the
thermodynamic limit $L \rightarrow \infty$,  will be analyzed in
the following subsection.

\subsection{Critical phenomena}
\label{sec:crit}

Following the finite-size scaling ansatz for continuous phase
transitions, equation~(\ref{eq:scal}), the leading critical
behavior of the interfacial adsorption, $W$, is given
by~\cite{Huse}
\begin{equation}
\label{eq:iaL} W(T_{\rm c}) \propto L^{a}
\end{equation}
and
\begin{equation}
\label{eq:iat} W(L \rightarrow \infty,t) \propto t^{b},
\end{equation}
with the critical exponents $a$ and $b$ being determined by the
bulk critical exponents $\beta$ and $\nu$~\cite{Huse}
\begin{equation}
\label{eq:ab} a = 1- \beta/\nu \;\; ; \;\; b=\beta -\nu.
\end{equation}
As stated before, the critical temperature, $T_{\rm c}$, is known
exactly for perfect and random Potts models on the square lattice,
equation~(\ref{eq:Critical point}).

These predictions have been confirmed quite reasonably in previous
Monte Carlo simulations for perfect Potts models with $q=3$ and
$4$. In the present study, we extend and refine the comparison by
considering randomness as well and by including corrections to
scaling. The (leading or effective) corrections may be cast in the
form, as usual,
\begin{equation}
\label{eq:iaL} W(T_{\rm c}) = W_0 L^{a} (1+c_0 L^{-x})
\end{equation}
and
\begin{equation}
\label{eq:iat} W(L \rightarrow \infty,t) = W_1 t^{b} (1+c_1t^{y})
\end{equation}
with $t=|T_{\rm c}-T|/T_{\rm c}$. In fact, the correction terms
play an important role when fitting the MC data for the lattice
sizes we simulated.

\begin{figure}
\resizebox{1 \columnwidth}{!}{\includegraphics{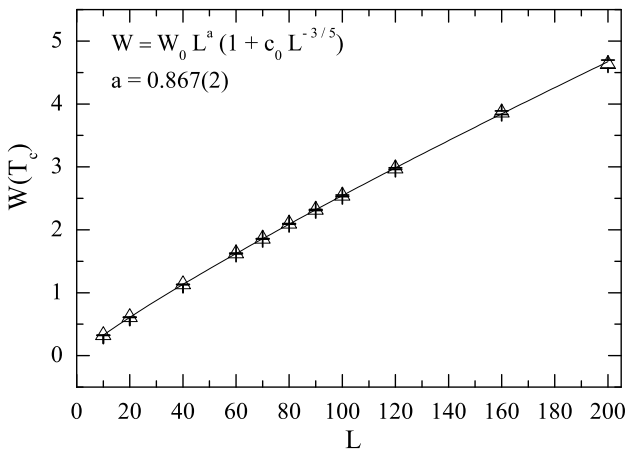}}
\caption{Finite-size scaling of the critical interfacial
adsorption of the pure 3-states Potts model, showing the
simulation data and the fitting curve.} \label{fig4}
\end{figure}

Let us first discuss results for the perfect 3-state Potts model.
Compared to previous simulations~\cite{Huse}, the statistics has
been improved significantly and larger lattice sizes have been
studied. The size dependence of the critical adsorption, $W(T_{\rm
c})$, is depicted in Figure~\ref{fig4}. The simulation data for
the complete lattice-size spectrum $L = 10 - 200$ have been
analyzed by fitting them to equation~(\ref{eq:iaL}). We have also
performed a fitting for the temperature dependence of the critical
adsorption of the form~(\ref{eq:iat}) for the largest linear size
studied, i.e., $L = 200$. The resulting estimates for the critical
exponents, obtained from good quality fittings with $\chi^{2}/{\rm
dof} \in \{0.6 - 1\}$, $a = 0.867(2)$ and $b = -0.728(6)$ agree
nicely within errors with the predicted exact values
$a=13/15=0.866\cdots$ (we remind the reader that $\beta / \nu =
2/15$) and $b = - 13 / 18 = -0.722\cdots$~\cite{Wu}. Note that we
fixed the leading correction-to-scaling exponents, $x$ and $y$, to
the predicted exact values, $x = 3 / 5$ and $y = 4 /
3$~\cite{DoFa,Priv,Shchur}. Likewise, the estimates for the
exponents, $x$ and $y$, are observed to agree with the predicted
values, when fixing the critical exponents $a$ and $b$ in the
fittings. Accordingly, the findings on the perfect 3-state Potts
model strongly support the correctness of the finite-size scaling
description~(\ref{eq:scal}).

\begin{figure}
\resizebox{1 \columnwidth}{!}{\includegraphics{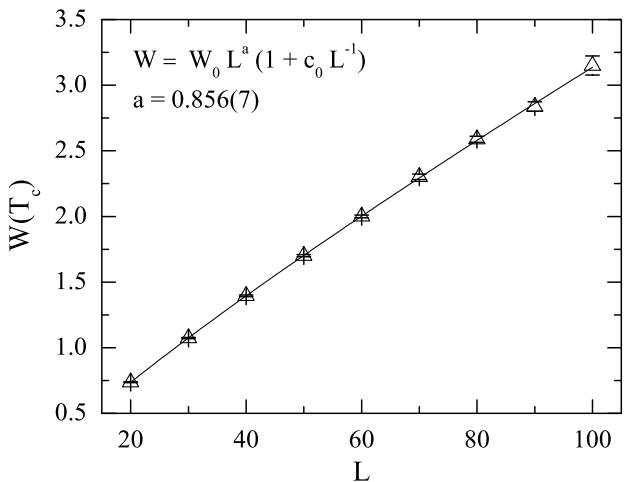}}
\caption{Finite-size scaling of the critical interfacial
adsorption data for the $q = 8$ and $r = 1 / 10$ diluted Potts
model, showing the simulation data and the fitting curve.}
\label{fig5}
\end{figure}

The main focus of the current study is on the dilute 8-state Potts
model. Following previous considerations on its bulk critical
properties, we set $r = J_2 / J_1 = 1 / 10$, where the randomness
dominated behavior is expected to show up already for moderate
lattice sizes~\cite{Berche,Chen,Jacob,Chat,Pala,Picco}. In
particular, we monitored, in our simulations, the size dependence
of the critical interfacial adsorption $W(T_{\rm c})$. As
discussed above, the standard errors stem from averaging over many
bond realizations. Numerical results are depicted in
Figure~\ref{fig5}. Fitting the MC data to equation~(\ref{eq:iaL})
with a correction-to-scaling exponent set to the value $x = -1$
which optimizes the fit, one obtains the value $a = 0.856(7)$.

For the above random case ($q=8$ and $r = 1 / 10$) and from
previous MC simulations and transfer-matrix
calculations~\cite{Berche,Chen,Jacob,Chat,Pala,Picco}, $\beta/\nu$
has been determined to be, approximately, $0.145 \pm 0.005$.
Actually, in the present simulations, we also recorded the size
dependence of the order parameter, for both types of fixed
boundary conditions, $m_{1:1}$ and $m_{1:2}$, vanishing, at
$T_{\rm c}$, as $\sim L^{-\beta/\nu}$. The resulting estimate for
the exponent, based on simple power-law fits to the magnetization
with systematically increasing the smallest lattice size, confirms
the previous findings. Actually, a linear extrapolation of the fit
exponent leads to the value $\beta / \nu = 0.145\pm 0.003$ for
$m_{1:1}$ as well as for $m_{1:2}$. Accordingly, we may safely
conclude for the random case as well $a = 1-\beta / \nu$, in
accordance with the finite-size scaling ansatz,
equation~(\ref{eq:scal}).

\begin{figure}
\resizebox{1 \columnwidth}{!}{\includegraphics{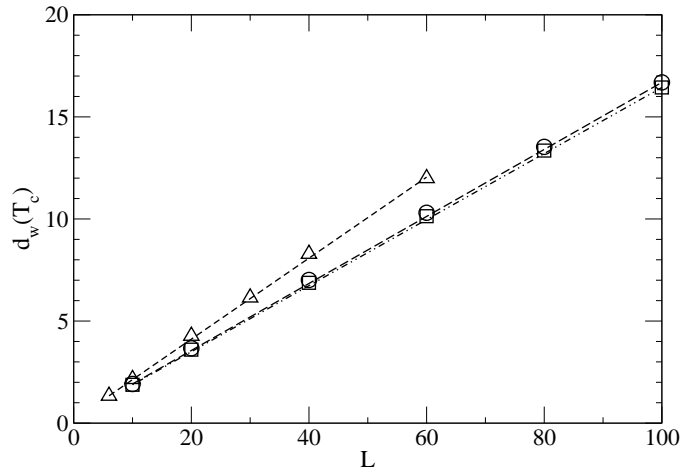}}
\caption{Size dependence of the critical spread of the adsorption
profiles $d_{\rm w}(T_{\rm c})$ in the perfect 3- (circles) and 4-
(squares) state as well as in the random, $r=1/10$, 8- (triangles)
state Potts models on square lattices with $L^2$ sites. Lines are
simple guides to the eye.} \label{fig6}
\end{figure}

We note in passing that our Monte Carlo data on the specific heat
demonstrate that the absolute value of the corresponding critical
exponent $\alpha/\nu$ is very small~\cite{Berche}. However, its
more accurate determination is beyond the scope of the present
study.

Finally, critical properties of the adsorption profile, $w_l$, and
its spread, $d_{\rm w}$, will be reported. As indicated by
Figure~\ref{fig3}, the spread $d_{\rm w}$ at $T_{\rm c}$ grows
when increasing the lattice size. We monitored the growth for the
perfect 3- and 4-state as well as for the dilute, with $r=1/10$,
3-, 4-, and 8-state Potts models on the square lattice. The
lattice size ranged from $L=6$ up to $L=100$. All data could be
fitted well to the ansatz
\begin{equation}
\label{eq:dw} d_{\rm w}(T_{\rm c},L) = d_{0} +d_{1} L,
\end{equation}
with the slope, $d_{1}$, depending rather strongly on the
randomness, but only weakly on $q$, as illustrated in
Figure~\ref{fig6}.

\section{Summary}
\label{sec:summary}

We performed extensive Monte Carlo simulations to study critical
interfacial properties in perfect and random ferromagnetic
$q$-state Potts models on the square lattice. In the dilute case,
there are two distinct, strong and weak, bonds connecting
neighboring sites. Both bonds are assumed to occur with the same
probability, leading to self-duality. Interfaces have been
introduced, by fixing the Potts variables at opposite sites in two
different states. The local Metropolis and the cluster-flip Wolff
algorithms have been used.

Randomness is found to affect, especially, the position of the
interface, the excess or interfacial adsorption, and the form of
the histograms resulting from the bond realizations.

Nevertheless, predictions of the isotropic finite-size scaling
description for the interfacial adsorption at continuous phase
transitions are observed to hold both for the perfect and random
cases, as has been exemplified for the pure 3-state and the dilute
8-state Potts models. In particular, critical exponents of the
interfacial adsorption are, indeed, determined by the bulk
critical exponents for the order parameter and the correlation
length. These bulk exponents depend on the number of Potts states,
$q$.

Moreover, we analyzed the adsorption profiles, describing the
spatial structure of the total interfacial adsorption. The spread
of the profiles seems to diverge at the phase transition in the
thermodynamic limit. For lattices with $L^{2}$ sites, we find a
linear increase with the linear dimension of the lattice for
perfect and random models, simulating the cases $q = 3$, $4$, and
$8$.

\begin{acknowledgement}
We would like to thank Alexei Brener for useful discussions in an
early stage of this project, as well as Bertrand Berche for a
helpful conversation. We thank Eren Metin El\c{c}i for his help 
in preparing Figure 1. A. Malakis acknowledges financial support
from Coventry University during a research visit at the Applied
Mathematics Research Centre, where part of this work has been
completed.
\end{acknowledgement}

\end{document}